\documentclass[twocolumn,pre,superscriptaddress]{revtex4}

\usepackage{dcolumn}
\usepackage{amsmath}

\usepackage{graphicx}
\usepackage{subfigure}
\usepackage{multirow}

\newcommand{\half}{\tfrac12}
\newcommand{\set}[1]{\lbrace#1\rbrace}

\newcommand{\etal}{{\it{}et~al.}}
\newcommand{\defn}{\textit}

\newcommand{\mat}{\mathbf}
\renewcommand{\vec}{\mathbf}
\newcommand{\argmax}{\mathop\mathrm{argmax}}

\newcommand\pin{p_\textrm{in}}
\newcommand\pout{p_\textrm{out}}
\newcommand\cin{c_\textrm{in}}
\newcommand\cout{c_\textrm{out}}

\begin{document}

\title{Structure and inference in annotated networks}

\author{M. E. J. Newman}
\affiliation{Department of Physics and Center for the Study of Complex
  Systems, University of Michigan, Ann Arbor, MI 48109, USA}
\affiliation{Santa Fe Institute, 1399 Hyde Park Road, Santa Fe, NM 87501}
\author{Aaron Clauset}
\affiliation{Santa Fe Institute, 1399 Hyde Park Road, Santa Fe, NM 87501}
\affiliation{Department of Computer Science and BioFrontiers Institute,
  University of Colorado, Boulder, CO 80309, USA}

\begin{abstract}
  For many networks of scientific interest we know both the connections of
  the network and information about the network nodes, such as the age or
  gender of individuals in a social network, geographic location of nodes
  in the Internet, or cellular function of nodes in a gene regulatory
  network.  Here we demonstrate how this ``metadata'' can be used to
  improve our analysis and understanding of network structure.  We focus in
  particular on the problem of community detection in networks and develop
  a mathematically principled approach that combines a network and its
  metadata to detect communities more accurately than can be done with
  either alone.  Crucially, the method does not assume that the metadata
  are correlated with the communities we are trying to find.  Instead the
  method learns whether a correlation exists and correctly uses or ignores
  the metadata depending on whether they contain useful information.  The
  learned correlations are also of interest in their own right, allowing us
  to make predictions about the community membership of nodes whose network
  connections are unknown.  We demonstrate our method on synthetic networks
  with known structure and on real-world networks, large and small, drawn
  from social, biological, and technological domains.
\end{abstract}

\maketitle

\section{Introduction}
Networks arise in many fields and provide a powerful and compact
representation of the internal structure of a wide range of complex
systems~\cite{Newman10}.  Examples include social networks of interactions
among people, technological and information networks such as the Internet
or the World Wide Web, and biological networks of molecules, cells, or
entire species.  The last two decades have witnessed rapid growth both in
the availability of network data and in the number and sophistication of
network analysis techniques.  Borrowing ideas from graph theory,
statistical physics, computer science, statistics, and other areas, network
analysis typically aims to characterize a network's structural features in
a way that sheds light on the behavior of the system the network describes.
Studies of social networks, for instance, might identify the most
influential or central individuals in a population.  Studies of road
networks can shed light on traffic flows or bottlenecks within a city or
country.  Studies of pathways in metabolic networks can lead to a more
complete understanding of the molecular machinery of the cell.

Most research in this area treats networks as objects of pure topology,
unadorned sets of nodes and their interactions.  Most network data,
however, are accompanied by annotations or \defn{metadata} that describe
properties of nodes such as a person's age, gender, or ethnicity in a
social network, feeding mode or body mass of species in a food web, data
capacity or physical location of nodes on the Internet, and so forth.
(There can be metadata data on the edges of a network as well as on the
nodes~\cite{AJC15}, but our focus here is on the node case.)  In this
paper, we consider how to extend the analysis of networks to directly
incorporate such metadata.  Our approach is based on methods of statistical
inference and can in principle be applied to a range of different network
analysis tasks.  Here, we focus specifically on one of the most widely
studied tasks, the community detection problem.  Community detection, also
called node clustering or classification, searches for a good division of a
network's nodes into groups or classes~\cite{Fortunato10}.  Typically, one
searches for \textit{assortative structure}, groupings of nodes such that
connections are denser within groups than between them.  This structure is
common in social networks, for example, where groups may correspond to sets
of friends or coworkers, but it also occurs in other cases, including
biological and ecological networks, the Web, transportation and
distribution networks, and others.  Less common, but no less important, is
\textit{disassortative structure}, in which network connections are sparser
within groups than between them, and mixtures of assortative and
disassortative structure can also occur, where different groups may have
varying propensities for within- or between-group connections.

In many cases, the groups identified by community detection correlate
meaningfully with other network properties or functions, such as
allegiances or personal interests in social
networks~\cite{Fortunato10,AG05} or biological function in metabolic
networks~\cite{HHJ03,GA05}.  Some recent research, however, has suggested
that these cases may be the exception rather than the
rule~\cite{YL12,HDF14}, an important point that we address later in this
paper.

A large number of methods have been proposed for detecting communities in
unannotated networks~\cite{Fortunato10}.  Among these, some of the most
powerful, both in terms of rigorously provable performance and of raw
speed, are those based on statistical inference.  Here we build on these
methods to incorporate node metadata into the community detection problem
in a principled and flexible manner.  The resulting methods have several
attractive features.  First, they can make use of metadata in arbitrary
format to improve the accuracy of community detection.  Second, and
crucially for our goals, they do not assume \textit{a~priori} that the
metadata correlate with the communities we seek to find.  Instead, they
detect and quantify the relationship between metadata and community, if one
exists, then exploit that relationship to improve the results.  Even if the
correlation is imperfect or noisy, the method can still use what
information is present to return improved results.  Conversely, if no
correlation exists the method will automatically ignore the metadata,
returning results based on network structure alone.

Third, our methods allow us to select between competing divisions of a
network.  Many networks have a number of different possible
divisions~\cite{GDC10}.  For example, a social network of acquaintances may
have meaningful divisions along lines of age, gender, race, religion,
language, politics, or many other variables.  By incorporating metadata
that correlate with a particular division of interest, we can favor that
division over others, steering the analysis in a desired direction.
(Approaches like this are sometimes referred to as ``supervised learning''
techniques, particularly in the statistics and machine learning
literature.)  Thus, if we are interested for instance in a division of a
social network along lines of age, and we have age data for some fraction
of the nodes, we can use those data to steer the algorithm toward
age-correlated divisions.  Even if the metadata are incomplete or noisy,
the algorithm can still use them to guide its analysis.  However, if we
hand the algorithm metadata that do not correlate with any good division of
the network, the method will decline to follow along blindly, and will
inform us that no good correlation exists.

Finally, the correlation between metadata and network structure learned by
the algorithm (if one exists) is interesting in its own right.  Once found,
it allows us to quantify the agreement between network communities and
metadata, and to predict community membership for nodes for which we lack
network data and have only metadata.  If we have learned, for example, that
age is a good predictor of social groupings, then we can make quantitative
predictions of group membership for individuals about whom we know their
age and nothing else.

A number of other researchers have investigated ways to incorporate
metadata into network analysis, though they have typically made stronger
assumptions about the relationship between metadata and
communities~\cite{BCMM15,YML13,BVR14}.  Perhaps closest to our approach are
the semi-supervised learning methods~\cite{Peel12,EM12,ZMZ14}, which treat
the case where we are given the exact community assignments of some
fraction of the nodes and the goal is to deduce the reminder.  A variant of
this approach is active learning, in which the community membership of some
nodes is given, but the known nodes are not specified \textit{a~priori},
being instead chosen by the algorithm itself as it
runs~\cite{YZRM11,Leng13}.  Another vein of research, somewhat further from
our approach, considers the case where we are told some pairs of nodes that
either definitely are or definitely are not in the same community, and then
assigns communities subject to these constraints~\cite{MGYF10,Zhang13}.

In the following sections we describe our method in detail, and apply it to
a selection of example networks.  We show that it recovers known
communities in benchmark data sets with higher accuracy than algorithms
based on network structure alone, that we can select between competing
community divisions in both real and synthetic tests, that the method is
able accurately to divine correlations between network structure and
metadata, or determine that no such correlation exists, and that learned
correlations between structure and metadata can be used to predict
community membership based on metadata alone.

\section{Methods}
\label{sec:methods}
Our method makes use of techniques of Bayesian statistical inference in
which we construct a generative network model possessing the specific
features we hope to find in our data, namely community structure and a
correlation between that structure and node metadata, then we fit the model
to an observed network plus accompanying metadata and the parameters of
the fit tell us about the structure of the network.

The model we use is a modified version of a \defn{stochastic block model}.
The original stochastic block model, proposed in 1983 by
Holland~\etal~\cite{HLL83}, is a simple model for generating random
networks with community structure in which nodes are divided among some
number of communities and edges are placed randomly and independently
between them with probabilities that depend only on the communities to
which the nodes belong.  We modify this model in two ways.  First,
following~\cite{KN11a}, we note that the standard stochastic block model
does poorly at mimicking the structure of networks with highly
heterogeneous degree sequences (which includes nearly all real-world
networks), and so we include a ``degree-correction'' term that matches node
degrees (i.e.,~the number of connections each node has) to those of the
observed data.  Second, we introduce a dependence on node metadata via a
set of prior probabilities.  The prior probability of a node belonging to a
particular community becomes a function of the metadata, and it is this
function that is learned by our algorithm in order to incorporate the
metadata into the calculation.

Consider an undirected network with $n$ nodes labeled by integers
$u=1\ldots n$, divided among~$k$ communities, and denote the community to
which node~$u$ belongs by~$s_u\in 1\ldots k$.  In the simplest case, we
consider metadata with a finite number~$K$ of discrete, unordered values
and we denote node~$u$'s metadata by $x_u\in 1\ldots K$.  The choice of
labels~$1\ldots K$ is arbitrary and does not imply an ordering for the
metadata or that the metadata are one-dimensional.  If a social network has
two-dimensional metadata describing both language and race, for example, we
simply encode each possible language/race combination as a different value
of~$x$: English/white, Spanish/white, English/black, and so forth.  If a
network has nodes that are missing metadata values, we just let ``missing''
be another metadata value.

Given metadata~$\vec{x} = \set{x_u}$ and degree~$\vec{d} = \set{d_u}$ for
all nodes, a network is generated from the model as follows.  First, each
node~$u$ is assigned to a community~$s$ with a probability depending on
$u$'s metadata~$x_{u}$.  The probability of assignment we
denote~$\gamma_{sx}$ for each combination $s,x$ of community and metadata,
so the full prior probability on community assignments is
$P(\vec{s}\,|\,\mat{\Gamma},\vec{x}) = \prod_i \gamma_{s_i,x_i}$, where
$\mat{\Gamma}$ denotes the $k\times K$ matrix of parameters~$\gamma_{sx}$.
(More complex forms of the prior are appropriate in other cases, as we will
see.)  Once every node has been assigned to a community, edges are placed
independently at random between nodes, with the probability of an edge
between nodes~$u$ and~$v$ being
\begin{equation}
p_{uv} = d_u d_v \theta_{s_u,s_v}\,.
\label{eq:pij}
\end{equation}
where $\theta_{st}$ are parameters that we specify, with $\theta_{st} =
\theta_{ts}$.  The factor $d_u d_v$ allows the model to fit arbitrary
degree sequences as described above.  Models of this kind have been found
to fit community structure in real networks well~\cite{KN11a}.

Community detection then consists of fitting the model to observed network
data using the method of maximum likelihood.  Given an observed network, we
define its adjacency matrix~$\mat{A}$ to be the $n\times n$ real symmetric
matrix with elements~$a_{uv} = 1$ if there is an edge between nodes~$u$
and~$v$ and 0 otherwise.  Then the probability, or likelihood, that this
network was generated by our model, given the parameters and metadata,
is
\begin{align}
P(\mat{A}\,|\,\mat{\Theta},\mat{\Gamma},\vec{x})
 &= \sum_{\vec{s}} P(\mat{A}\,|\,\mat{\Theta},\vec{s})
    P(\vec{s}\,|\,\mat{\Gamma},\vec{x}) \nonumber\\
 &= \sum_{\vec{s}} \prod_{u<v} p_{uv}^{a_{uv}} (1-p_{uv})^{1-a_{uv}}
    \prod_u \gamma_{s_u,x_u} \, ,
\end{align}
where $\mat{\Theta}$ is the $k\times k$ matrix with
elements~$\theta_{st}$ and the sum is over all possible community
assignments~$\vec{s}$.

Fitting the model involves maximizing this likelihood with respect
to~$\mat{\Theta}$ and~$\mat{\Gamma}$ to determine the most likely values of
the parameters, which we do using an expectation--maximization (EM)
algorithm.  A full derivation of the algorithm is given in
Appendix~\ref{app:em}, but the central result is that the optimal parameter
values are
\begin{equation}
\theta_{st} = {\sum_{uv} a_{uv} q_{st}^{uv}\over
               \sum_u d_u q_s^u \sum_v d_v q_t^v} \, , \quad
\gamma_{sx} = {\sum_u \delta_{x,x_u} q_s^u\over\sum_u \delta_{x,x_u}} \, ,
\label{eq:thetagamma}
\end{equation}
where $\delta_{xy}$~is the Kronecker delta and
\begin{equation}
q_s^u = \sum_{\vec{s}} q(\vec{s}) \delta_{s_i,s} \, ,\qquad
q_{st}^{uv} = \sum_{\vec{s}} q(\vec{s}) \delta_{s_u,s} \delta_{s_v,t} \, ,
\label{eq:marginals}
\end{equation}
with
\begin{equation}
q(\vec{s}) = {P(\mat{A}\,|\,\mat{\Theta},\vec{s})
  P(\vec{s}\,|\,\mat{\Gamma},\vec{x})\over
  \sum_{\vec{s}} P(\mat{A}\,|\,\mat{\Theta},\vec{s})
               P(\vec{s}\,|\,\mat{\Gamma},\vec{x})} 
  = P(\vec{s}\,|\,\mat{A},\mat{\Theta},\mat{\Gamma},\vec{x}).
\label{eq:qz}
\end{equation}
Physically, $q_s^u$~is the marginal posterior probability that node~$u$
belongs to community~$s$ and $q_{st}^{uv}$ is the joint probability that
nodes $u$ and~$v$ belong to $s$ and $t$ respectively.  Normally, in fact,
$q_r^u$~is the object of primary interest in the calculation, as it tells
us to which group each node belongs, i.e.,~it tells us the optimal division
of the network into communities.  The prior probabilities~$\gamma_{sx}$ are
also of interest, since they tell us how and to what extent the metadata
are correlated with the communities, a point discussed further in
Section~\ref{sec:results}.

Computationally, the most demanding part of the EM algorithm is calculating
the sum in the denominator of Eq.~\eqref{eq:qz}, which has an exponentially
large number of terms, making its direct evaluation intractable on all but
the smallest of networks.  Traditionally one gets around this problem by
approximating the full distribution~$q(\vec{s})$ by Monte Carlo importance
sampling.  Here, we instead use a recently proposed alternative method
based on \defn{belief propagation}~\cite{DKMZ11a}, which is significantly
faster and fast enough in practice for applications to very large networks.
(In separate work we have successfully applied the method of this paper to
a network of over 1.4 million nodes.)

We also consider cases in which the metadata are ordered and potentially
continuous variables, such as age or income in a social network, which
require a different algorithm.  The prior probability $P(s\,|\,x)$ of
belonging to community~$s$ given metadata value~$x$ becomes a continuous
function of~$x$, which we write as an expansion in a set of basis
functions~$B_j(x)$, parametrized by the coefficients~$\gamma_{sj}$ of the
expansion:\ $P(s\,|\,x) = \sum_{j=0}^N \gamma_{sj} B_j(x)$.  The result for
the optimal value of~$\theta_{st}$ is still given by
Eq.~\eqref{eq:thetagamma}, but the optimal values of the~$\gamma_{sj}$ are
given by the solution of the equations
\begin{equation}
\gamma_{sj} = {\sum_u q_s^u Q_j^{su}\over\sum_{tu} q_t^u Q_j^{tu}} \, ,
\qquad
Q_j^{su} = {\gamma_{sj} B_j(x_u)\over\sum_k \gamma_{sk} B_k(x_u)} \, .
\end{equation}
A full derivation is given in Appendix~\ref{app:em}.  These equations can
be conveniently and rapidly solved by simple iteration, starting with the
current best estimate of~$\gamma_{sj}$ and alternating between the
equations until convergence is achieved.  In our implementation, we use
Bernstein polynomials as the basis functions, although other choices are
possible.

\section{Results}
\label{sec:results}
We have applied the method to a range of example networks, including
computer-generated benchmarks that test its ability to detect known
structure, as well as a variety of real-world networks.

\subsection{Synthetic networks}\ Our first tests are on computer-generated
(``synthetic'') networks that have known community structure embedded
within them.  These networks were created using the standard stochastic
block model, in which nodes are assigned to groups, then edges are placed
between them independently with probabilities that are a function of group
membership only~\cite{HLL83,KN11a}.  After the networks are created, we
generate discrete-valued node metadata at random that match the true
community assignments of nodes a given fraction of the time and are chosen
randomly from the non-matching values otherwise.  This allows us to control
the extent to which the metadata correlate with the community structure and
hence test the algorithm's ability to make use of metadata of varying
quality.

\begin{figure}[t]
\begin{center}
\includegraphics[width=\columnwidth]{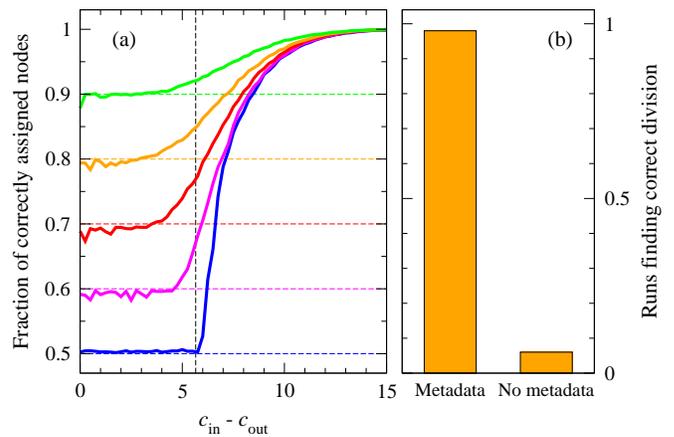}
\end{center}
\caption{Tests on synthetic benchmark networks with $n=10\,000$ nodes.
  (a)~Fraction of correctly assigned nodes for networks with two planted
  communities with mean degree $c=8$, as a function of the difference
  between the numbers of
  within- and between-group connections.  The five curves show results for
  networks with a match between metadata and planted communities on a
  fraction 0.5, 0.6, 0.7, 0.8, and~0.9 of nodes (bottom to top).  The
  vertical dashed line indicates the theoretical detectability threshold,
  below which no algorithm without metadata can detect the communities.
  (b)~Fraction of 100 four-group test networks where the algorithm selects
  a particular 2-way division, out of several competing possibilities, with
  and without the help of metadata that are weakly correlated with the
  desired division.  A run is considered to find the correct division if
  the fraction of correctly classified nodes exceeds~85\%.  Network
  parameters are $\cout=4$ and~$\cin=20$.
\label{fig:sbm}}
\end{figure}

Figure~\ref{fig:sbm}a shows results for a set of such networks with two
communities of equal size, with edge probabilities $\pin = \cin/n$ and
$\pout = \cout/n$ for within-group and between-group edges, respectively.
When $\cin$ is much greater than $\cout$ the communities are easy to detect
from network structure alone, but as $\cin$ approaches $\cout$ the
structure becomes weaker and harder to detect.  Each curve in the figure
shows the fraction of nodes that are classified into their correct groups
by our algorithm, as we vary the strength of the community structure,
measured by the difference $\cin-\cout$.  Individual curves show results
for different levels of correlation between communities and metadata.

When metadata and community agree for exactly half of the nodes (bottom
curve) there is no correlation between the two and the metadata cannot help
in community detection.  It thus comes as no surprise that this curve shows
the lowest success rate.  For the higher levels of correlation the metadata
contain useful information and the algorithm's performance improves
accordingly.

Examining the figure, a clear pattern emerges.  For large $\cin-\cout$ the
network contains strong community structure and the algorithm reliably
classifies essentially all nodes into the correct groups, as we would
expect of any effective algorithm.  As the structure weakens the fraction
of correct nodes declines, but it remains higher in all the cases where the
metadata are useful than in the lowest curve where they are not.  Moreover,
the algorithm's success rate appears to improve monotonically with the
level of correlation between metadata and communities.

When there are no metadata, it is known that the EM algorithm gives optimal
answers to the community detection problem in the sense that no other
algorithm will classify a higher fraction of nodes correctly on
average~\cite{DKMZ11a}.  The fact that our algorithm does better when there
are metadata thus implies that \textit{the algorithm with metadata does
  better than any possible algorithm without metadata.}

Furthermore, it has previously been shown that below the so-called
\textit{detectability threshold}, which occurs at $\cin-\cout =
\sqrt{2(\cin+\cout)}$ (indicated by the vertical dashed line in the figure,
and aligning with the sharp transition in the bottom curve), community
structure becomes so weak as to be undetectable by any algorithm that
relies on network structure alone~\cite{DKMZ11a,MNS13}.  Well below this
threshold, however, our algorithm still correctly classifies a fraction of
the nodes roughly equal to the fraction of metadata that match the
communities, meaning that the algorithm does better with metadata than
without it even below the threshold.  One can understand this result
theoretically by observing that $\gamma_{sx} = \delta_{sx}$ is a fixed
point of Eqs.~\eqref{eq:thetagamma} to~\eqref{eq:qz}, so that assigning
each node~$u$ to the group indicated by its metadata~$x_u$ is always a
solution.  Figure~\ref{fig:sbm}a also shows that the fraction of correctly
classified nodes beats this baseline level for values of $\cin-\cout$
somewhat below the threshold, suggesting that the use of the metadata
shifts the threshold downward or perhaps eliminates it altogether.

In short, our method automatically combines the available information from
network structure and metadata to do a better job of community detection
than any algorithm based on network structure alone.  And when either the
network or the metadata contain no information about community structure
the algorithm correctly ignores them and returns an estimate based only on
the other.

Figure~\ref{fig:sbm}b shows a different synthetic test, of the algorithm's
ability to select between competing divisions of a network.  In this test,
networks were generated with four equally sized communities but the
algorithm was tasked with finding a division into just two communities.
There are eight ways of dividing such a network in two if we are to keep
the four underlying groups undivided.  We imagine a situation in which we
are interested in finding a particular one out of these eight.  A
conventional community detection algorithm may find a reasonable division
of the network, but there is no guarantee it will find the ``correct''
one---some fraction of the time we can expect it to find one of the
competing divisions.  But if the algorithm is given a set of metadata that
correlate with the division of interest, even if the correlation is poor,
the likelihood of that division will be increased relative to the others
and it will become favored.

In our tests the desired division was one that places two of the underlying
four groups in one community and the remaining two in the other.
Two-valued metadata were generated that agree with this division 65\% of
the time, a relatively weak level of correlation, not far above the 50\% of
completely uncorrelated data.  Nonetheless, as shown in
Figure~\ref{fig:sbm}b, this is enough for the algorithm to reliably find
the correct division of the network in almost every case---98\% of the time
in our tests.  Without the metadata, by contrast, we succeed only 6\% of
the time.  Some practical applications of this ability to select among
competing divisions are given in the next section.

\subsection{Real-world networks}\ In this section we describe applications
to three real-world networks, drawn from social, biological, and
technological domains respectively.  Two further applications are given in
Appendix~\ref{app:examples}.

\smallskip \noindent \textit{School friendships}:\ For our first
application we analyze a network of school students, drawn from the US
National Longitudinal Study of Adolescent Health~\cite{Note1}.  The network
represents patterns of friendship, established by survey, among the 795
students in a medium-sized American high school (US grades 9 to 12, ages 14
to 18 years) and its feeder middle school (grades 7 and 8, ages 12 to 14
years).

Given that this network combines middle and high schools, it comes as no
surprise that there is a clear division (previously documented) into two
network communities corresponding roughly to the two
schools~\cite{Moody01}.  Previous work, however, has also shown the
presence of divisions by ethnicity.  Our method allows us to select between
divisions by using metadata that correlate with the one in which we are
interested.

\begin{figure}[t]
\begin{center}
\includegraphics[width=8cm,clip=true]{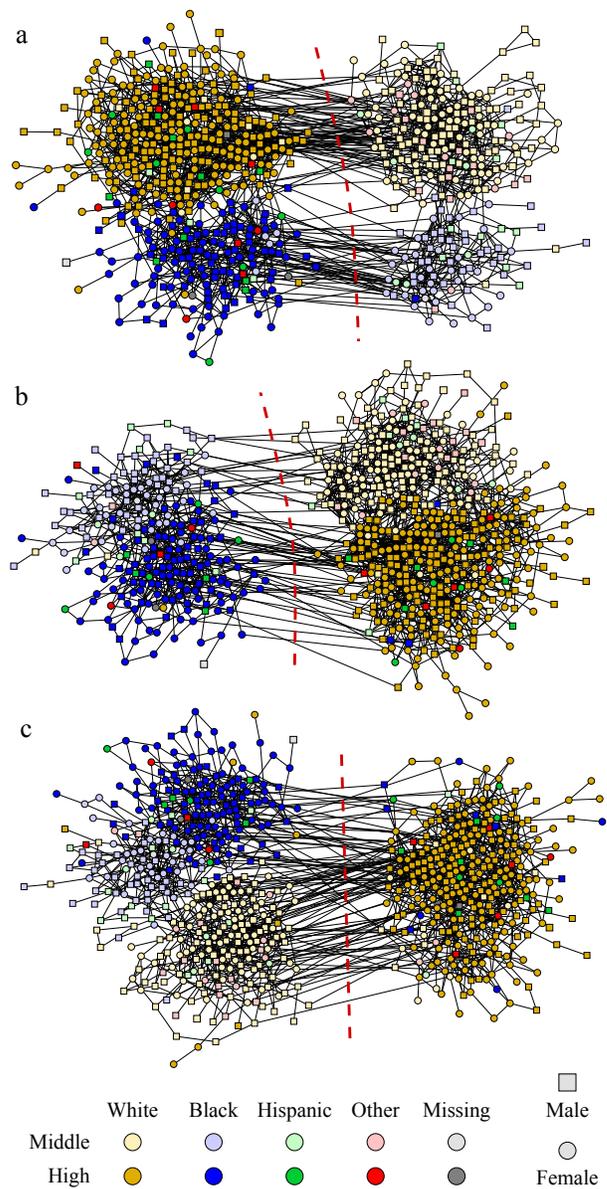}
\end{center}
\caption{Three divisions of a school friendship network, using as metadata
  (a)~school grade, (b)~ethnicity, and (c)~gender.\label{fig:addhealth}}
\end{figure}

Figure~\ref{fig:addhealth} shows the results of applying our algorithm to
the network three times.  Each time, we asked the algorithm to divide the
network into two communities. In Fig.~\ref{fig:addhealth}a, we used the six
school grades as metadata and the algorithm readily identifies a division
into grades 7 and 8 on the one hand and grades 9 to 12 on the
other---i.e.,~the division onto middle school and high school.  In
Fig.~\ref{fig:addhealth}b, by contrast, we used the students'
self-identified ethnicity as metadata, which in this data set takes one of
four values:\ white, black, hispanic, or other (plus a small number of
nodes with missing data).  Now the algorithm finds a completely different
division into two groups, one group consisting principally of black
students and one of white.  (The small number of remaining students are
distributed roughly evenly between the groups.)

One might be concerned that in these examples the algorithm is mainly
following the metadata to determine community memberships, and ignoring the
network structure.  To test for this possibility, we performed a third
analysis, using gender as metadata.  When we do this, as shown in
Fig.~\ref{fig:addhealth}c, the algorithm does not find a division into male
and female groups.  Instead, it finds a new division that is a hybrid of
the grade and ethnicity divisions (white high-school students in one group
and everyone else in the other).  That is, the algorithm has ignored the
gender metadata, because there was no good network division that correlated
with it, and instead found a division based on the network structure alone.
The algorithm makes use of the metadata only when doing so improves the
quality of the network division, in the sense of increasing the value of
the likelihood.

The extent to which the communities found by our algorithm match the
metadata (or any other ``ground truth'' variable) can be quantified by
calculating a normalized mutual information (NMI)~\cite{DDDA05,MGH11}.  NMI
ranges in value from 0 when the metadata are uninformative about the
communities to 1 when the metadata specify the communities completely.
(See the Supplemental Information for a detailed definition and discussion
of normalized mutual information.)  The divisions shown in
Fig.~\ref{fig:addhealth}a and~\ref{fig:addhealth}b have NMI scores of 0.881
and 0.820 respectively, indicating that the metadata are strongly though
not perfectly correlated with community membership.  By contrast, the
division in Fig.~\ref{fig:addhealth}c, where gender was used as metadata,
has an NMI score of 0.003, indicating that the metadata contain essentially
zero information about the communities.

\begin{figure}[t]
\begin{center}
\includegraphics[width=\columnwidth]{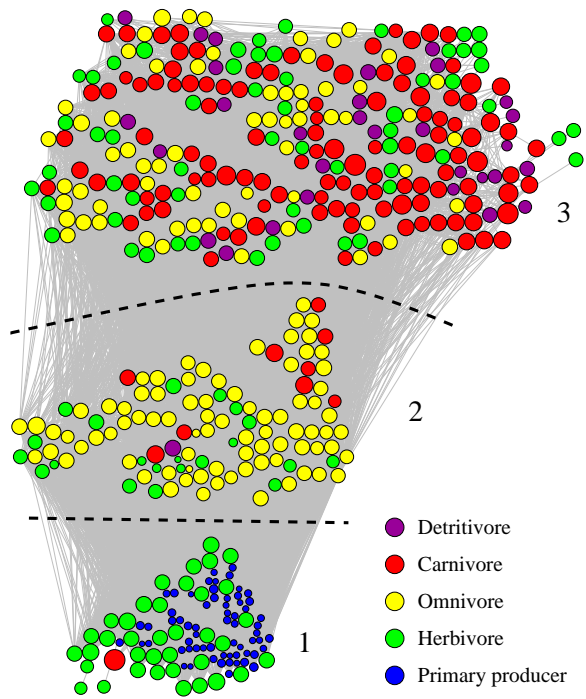}
\end{center}
\vspace{-2mm}
\caption{Three-way decomposition of the marine food web described in the
  text, with the logarithm of mean body mass used as metadata.  Node sizes
  are proportional to log-mass and colors indicate species role within the
  ecosystem.\label{fig:weddell}}
\end{figure}

\smallskip \noindent \textit{Predator-prey interactions}:\ Our next
application is one with ordered metadata of the kind described in
Section~\ref{sec:methods}.  The network in this case is an ecological one,
a food web of predator--prey interactions between 488 marine species living
in the Weddell Sea, a large bay off the coast of Antarctica.  A~number of
different metadata are available for these species, including feeding mode
(deposit feeder, suspension feeder, scavenger, etc.), zone within the ocean
(benthic, pelagic, etc.), and others.  In our analysis, however, we focus
on one in particular, the average adult body mass.  Body masses of species
in this ecosystem have a wide range, from microorganisms weighing nanograms
or less to hundreds of tonnes for the largest whales.  Conventionally in
such cases one often works with the logarithm of mass, which makes the
range more manageable, and we do so here.  Then we perform $k$-way
community decompositions using this log-mass as metadata, for various
values of~$k$.

Figure~\ref{fig:weddell} shows the results for $k\!=\!3$.  Nodes are
colored according to their role in the ecosystem---carnivores, herbivores,
primary producers, and so forth.  The division found by the algorithm
appears to match these roles quite closely, with one group composed almost
entirely of primary producers and herbivores, one of omnivores, and a third
that contains most of the carnivores.  Node sizes in the figure are
proportional to log-mass, which increases as we go up the figure,
indicating that the algorithm has recovered from the network structure the
well-known correlation between body mass and ecosystem
role~\cite{woodward05}.  This point is further emphasized by the values of
the prior probabilities of membership in the three groups as a function of
body mass (see Fig.~S1 in the Supplemental Information), which show that
low-mass organisms are overwhelmingly likely to be in the first group, and
high-mass ones in the third group.  Organisms of intermediate mass have a
broader distribution, but are particularly concentrated in the second
group.

The prior probabilities are also of interest in their own right.  If, for
instance, we were to learn of a new species, previously unrepresented in
our food-web data set, then even without knowing its pattern of network
connections we can make a statement about its probability of belonging to
each of the communities, as well as its probability of interaction with
other species, so long as we know its body mass. For instance, a low body
mass of $10^{-12}\,$g would put a species with high probability in group~1
in Fig.~\ref{fig:weddell}, meaning it is almost certainly a primary
producer or a herbivore, with the interaction patterns that implies.

\smallskip \noindent \textit{Internet graph}:\ Community detection is
widely studied precisely because it is believed that network communities
are correlated with network function.  More specifically, it is commonly
assumed that communities correlate with some underlying functional
variable, which may or may not be observed.  This assumption, however, has
been challenged by recent work that compared communities in real-world
networks against ``ground truth'' metadata variables and found little
correlation between the two~\cite{YL12,HDF14}.  This is a striking
discovery, but there is a caveat.  As we have seen, there are often
multiple meaningful community divisions of a network (as in the school
friendship network of Fig.~\ref{fig:addhealth}, for example), and the fact
that one division is uncorrelated with a given metadata variable does not
rule out the possibility that another could be.

Our third real-world example application illustrates these issues using one
of the same networks studied in Ref.~\cite{HDF14}, a $46\,676$-node
representation of the peering structure of the Internet at the level of
autonomous systems.  The ``ground truth'' variable for this network is the
country in which each autonomous system is located.  The analysis
of~\cite{HDF14} found there to be little correlation between community
structure and countries.

We first analyze the network without metadata, performing a traditional
``blind'' community division, into five groups using the standard EM
algorithm.  We then repeat the analysis using the algorithm of this paper
with the countries as metadata.  Recall that, in doing this, we do not
force the algorithm to find a community division that aligns with the
metadata if no such division exists, but if a division does exist it will
be favored over competing divisions that do not align with the metadata.
There are 173 distinct countries in the data set, a significantly larger
number of metadata values than for any of the other networks we have
considered, but by no means beyond the capabilities of our method.

As before, we assess the results using the normalized mutual information.
If indeed there are many competing divisions of the network, only some of
which correlate with the particular metadata we are given, then we would
expect our blind analysis to return a range of NMI values on different
runs, some low and (maybe) some higher.  This is indeed what we see, with
the NMI in our calculations ranging from a high of 0.626 to a relatively
low~0.398, the latter being in agreement with results quoted
in~\cite{HDF14}.  Conversely, when the algorithm of this paper is applied
with countries as metadata, we find an NMI score significantly higher than
any of these figures, at 0.870, which would conventionally be interpreted
as an indication of strong correlation.

These results emphasize that an apparent lack of correlation between
network communities and metadata could be the result of the presence of
competing network divisions, some of which are correlated with the
particular metadata we have in hand while others are not.  The algorithm of
this paper allows us to select among divisions and hence find ones that
correlate with the variable of interest.

\section{Conclusions}
In this paper we have described a technique for directly incorporating
annotations or ``metadata'' into the analysis of networks.  We have focused
on the problem of community detection, although the methods we describe
could in principle be applied to other analyses.  We have shown that the
incorporation of metadata improves the accuracy of community detection in
controlled studies on benchmark networks and also allows us to select among
competing community divisions in the same network, a common feature of
practical network data sets.  Our method is able to infer the level of
correlation between metadata and network structure, and will automatically
use or ignore the metadata as appropriate based on this inference.  We have
demonstrated applications of the method to a variety of data sets,
including social, biological, and technological networks, finding improved
results and flexibility of analysis compared to methods that do not use
metadata.

There are a number of possible extensions of this work.  At the simplest
level one could include more complex metadata types, such as combinations
of discrete and continuous variables, or vector variables such as spatial
coordinates.  Metadata could also be incorporated into methods for
detecting other structure types, such as hierarchy, core--periphery
structure, rankings, or latent-space structure.  And the resulting fits
could form the starting point for a variety of additional applications,
such as the prediction of missing links or missing metadata in incomplete
data sets.  These and other possibilities we leave for future work.

\begin{acknowledgments}
  The authors thank Cristopher Moore, Daniel Larremore, and Leto Peel for useful
  conversations and Darko Hric, Richard Darst, and Santo Fortunato for
  sharing the Internet data set and taking the time to explain it to us.
  This research was funded in part by the US National Science Foundation
  under grants DMS--1107796 and DMS--1407207 (MEJN) and IIS--1452718 (AC).
\end{acknowledgments}

\appendix

\section{EM algorithm}
\label{app:em}
In this appendix we present the derivation of the expectation--maximization
(EM) algorithm used to fit our model to empirical network data.

\subsection{Unordered data}
Given a network, represented by its adjacency matrix~$\mat{A}$, plus the
accompanying vector of metadata~$\vec{x}$, our goal is to determine the
values of the parameter matrices $\boldsymbol{\Theta},\boldsymbol{\Gamma}$
that maximize the likelihood of the network
\begin{equation}
P(\mat{A}|\boldsymbol{\Theta},\boldsymbol{\Gamma},\vec{x})
 = \sum_{\vec{s}} P(\mat{A}|\boldsymbol{\Theta},\vec{s})
    P(\vec{s}|\boldsymbol{\Gamma},\vec{x}),
\label{eq:likelihood}
\end{equation}
where
\begin{equation}
P(\mat{A}|\boldsymbol{\Theta},\vec{s})
 = \sum_{\vec{s}} \prod_{u<v} p_{uv}^{a_{uv}} (1-p_{uv})^{1-a_{uv}}
\label{eq:Pats}
\end{equation}
and
\begin{equation}
P(\vec{s}|\boldsymbol{\Gamma},\vec{x}) = \prod_u \gamma_{s_u,x_u},
\label{eq:Psgx}
\end{equation}
with
\begin{equation}
p_{uv} = d_u d_v \theta_{s_u,s_v},
\end{equation}
and $d_u$ being the degree of node~$u$.  Typically, rather than
maximizing~\eqref{eq:likelihood} itself, we maximize instead its logarithm,
\begin{equation}
\log P(\mat{A}|\boldsymbol{\Theta},\boldsymbol{\Gamma},\vec{x})
 = \log \sum_{\vec{s}} P(\mat{A}|\boldsymbol{\Theta},\vec{s})
    P(\vec{s}|\boldsymbol{\Gamma},\vec{x}),
\label{eq:ll}
\end{equation}
which gives the same answer for $\boldsymbol{\Theta}$ and
$\boldsymbol{\Gamma}$ but is often more convenient.

The most obvious approach for performing the maximization would be simply
to differentiate with respect to the parameters, set the result to zero,
and solve the resulting equations.  This, however, produces a complex set
of implicit equations that have no easy solution.  Instead, therefore, we
make use of Jensen's inequality, which says that for any set of positive
quantities~$x_i$ the log of their sum obeys
\begin{equation}
\log \sum_i x_i \ge \sum_i q_i \log {x_i\over q_i},
\label{eq:jensen}
\end{equation}
where $q_i$ is any correctly normalized probability distribution such that
$\sum_i q_i = 1$.  Note that the exact equality is recovered by the
particular choice
\begin{equation}
q_i = {x_i\over\sum_i x_i}.
\label{eq:exact}
\end{equation}

Applying Jensen's inequality to Eq.~\eqref{eq:ll}, we find that
\begin{align}
\log P(\mat{A}|\boldsymbol{\Theta},\boldsymbol{\Gamma},\vec{x})
   &\ge \sum_{\vec{s}} q(\vec{s})
    \log {P(\mat{A}|\boldsymbol{\Theta},\vec{s})
          P(\vec{s}|\boldsymbol{\Gamma},\vec{x})\over q(\vec{s})} \nonumber\\
  &\hspace{-6em}{} = \sum_{\vec{s}} q(\vec{s})
     \log P(\mat{A}|\boldsymbol{\Theta},\vec{s}) +
     \sum_{\vec{s}} q(\vec{s}) \log P(\vec{s}|\boldsymbol{\Gamma},\vec{x})
     \nonumber\\
  &\hspace{-5em}{} - \sum_{\vec{s}} q(\vec{s}) \log q(\vec{s}),
\label{eq:ineq}
\end{align}
where $q(\vec{s})$ is any distribution over community assignments~$\vec{s}$
such that $\sum_{\vec{s}} q(\vec{s}) = 1$.  The maximum of the right-hand
side of this inequality with respect to possible choices of the
distribution~$q(\vec{s})$ coincides with the exact equality, which,
following Eq.~\eqref{eq:exact}, is when
\begin{equation}
q(\vec{s}) = {P(\mat{A}|\boldsymbol{\Theta},\vec{s})
              P(\vec{s}|\boldsymbol{\Gamma},\vec{x})\over
  \sum_{\vec{s}} P(\mat{A}|\boldsymbol{\Theta},\vec{s})
                 P(\vec{s}|\boldsymbol{\Gamma},\vec{x})}.
\label{eq:qs}
\end{equation}
Thus the maximization of the left-hand side of~\eqref{eq:ineq} with respect
to $\boldsymbol{\Theta},\boldsymbol{\Gamma}$ to give the optimal values of
the parameters is equivalent to a maximization of the right-hand side both
with respect to $q(\vec{s})$ (which makes it equal to the left-hand side)
and with respect to~$\boldsymbol{\Theta},\boldsymbol{\Gamma}$.  A simple
algorithm for performing such a double maximization is to repeatedly
maximize with respect to first~$q(\vec{s})$ and then
$\boldsymbol{\Theta},\boldsymbol{\Gamma}$ until we converge to an answer.
In other words:
\begin{enumerate}
\item Make an initial guess about the parameter values and use
them to calculate the optimal~$q(\vec{s})$ from Eq.~\eqref{eq:qs}.
\item Using that value, maximize the right-hand side of~\eqref{eq:ineq}
  with respect to the parameters, while holding~$q(\vec{s})$ constant.
\item Repeat from step 1 until convergence is achieved.
\end{enumerate}
Step~2 can be performed by differentiating with $q(\vec{s})$ fixed and
subject to the normalization constraint $\sum_s \gamma_{sx} = 1$ for
all~$x$.  Performing the derivatives and assuming that the network is large
and sparse so that $p_{uv}$ is small, we find to leading order in small
quantities that
\begin{equation}
\theta_{st} = {\sum_{uv} a_{uv} q_{st}^{uv}\over
               \sum_{uv} d_u d_v q_{st}^{uv}}, \qquad
\gamma_{sx} = {\sum_u \delta_{x,x_u} q_s^u\over\sum_u \delta_{x,x_u}},
\label{eq:appthetagamma}
\end{equation}
where
\begin{equation}
q_s^u = \sum_{\vec{s}} q(\vec{s}) \delta_{s_i,s},\quad
q_{st}^{uv} = \sum_{\vec{s}} q(\vec{s}) \delta_{s_u,s} \delta_{s_v,t}.
\end{equation}
In addition, for a large sparse network, the community assignments of
distant nodes will be uncorrelated and hence we can
write~$q_{st}^{uv}\simeq q_s^u q_t^v$ in the denominator
of~\eqref{eq:appthetagamma} to get
\begin{equation}
\theta_{st} = {\sum_{uv} a_{uv} q_{st}^{uv}\over
               \sum_u d_u q_s^u \sum_v d_v q_t^v},
\end{equation}
which reduces the denominator sums from $n^2$ terms to only~$n$ and
considerably speeds the calculation.  (We cannot make the same
factorization in the numerator, since the terms in the numerator involve
$q_{st}^{uv}$ on adjacent nodes~$u,v$ only, so the nodes are not distant
from one another.)

Equation~\eqref{eq:qs} tells us that once the iteration converges, the
value of~$q(\vec{s})$ is
\begin{align}
q(\vec{s}) &= {P(\mat{A}|\boldsymbol{\Theta},\vec{s})
              P(\vec{s}|\boldsymbol{\Gamma},\vec{x})\over
  \sum_{\vec{s}} P(\mat{A}|\boldsymbol{\Theta},\vec{s})
                 P(\vec{s}|\boldsymbol{\Gamma},\vec{x})}
  = {P(\mat{A},\vec{s}|\boldsymbol{\Theta},\boldsymbol{\Gamma},\vec{x})\over
     P(\mat{A}|\boldsymbol{\Theta},\boldsymbol{\Gamma},\vec{x})}
     \nonumber\\
 &= P(\vec{s}|\mat{A},\boldsymbol{\Theta},\boldsymbol{\Gamma},\vec{x}).
\end{align}
In other words $q(\vec{s})$ is the posterior distribution over community
assignments~$\vec{s}$, the probability of an assignment~$\vec{s}$ given the
inputs~$\mat{A}$, $\boldsymbol{\Theta}$, $\boldsymbol{\Gamma}$,
and~$\vec{x}$.

\subsection{Final likelihood value}
The EM algorithm always converges to a maximum of the likelihood but is not
guaranteed to converge to the global maximum---it is possible for there to
be one or more local maxima as well.  To get around this problem we
normally run the algorithm repeatedly with different random initial guesses
for the parameters and from the results choose the one that finds the
highest likelihood value.  In the calculations presented in this paper we
did at least ten such ``random restarts'' for each network.

To determine which run has the highest final value of the likelihood we
calculate the likelihood from the right-hand side of~\eqref{eq:ineq} using
$P(\mat{A}|\boldsymbol{\Theta},\vec{s})$ and
$P(\vec{s}|\boldsymbol{\Gamma},\vec{x})$ as in Eqs.~\eqref{eq:Pats}
and~\eqref{eq:Psgx}, the final fitted values of the
parameters~$\boldsymbol{\Theta}$ and $\boldsymbol{\Gamma}$ from the EM
algorithm, and $q(\vec{s})$ as in Eq.~\eqref{eq:qs}.  (As we have said, the
right-hand side of~\eqref{eq:ineq} becomes equal to the left, and hence
equal to the true log-likelihood, when $q(\vec{s})$ is given the value in
Eq.~\eqref{eq:qs}.)

\begin{widetext}
  Putting it all together, our expression for the log-likelihood is
\begin{align}
\log P(\mat{A}|\boldsymbol{\Theta},\boldsymbol{\Gamma},\vec{x})
  &= \sum_{\vec{s}} q(\vec{s}) \sum_{u<v} \bigl[ a_{uv} \log(d_u d_v
  \theta_{s_u,s_v}) + (1-a_{uv}) \log(1 - d_u d_v \theta_{s_u,s_v}) \bigr]
  \nonumber\\
  &\hspace{15em}{} + \sum_{\vec{s}} q(\vec{s}) \sum_u \log \gamma_{s_u,x_u} 
  - \sum_{\vec{s}} q(\vec{s}) \log q(\vec{s}).
\label{eq:ll1}
\end{align}
Neglecting terms beyond first order in small quantities, the first sum can
be rewritten as
\begin{align}
&\half \sum_{uv} \sum_{st} \bigl[ q_{st}^{uv} a_{uv} (\log d_u + \log d_v
  + \log \theta_{st} ) - q_{st}^{uv} d_u d_v \theta_{st} \bigr] \nonumber\\
  &\hspace{4em}{} = \half \biggl[ \sum_u d_u \log d_u + \sum_v d_v \log d_v 
  + \sum_{st} \log \theta_{st} \sum_{uv} a_{uv} q_{st}^{uv}
  - \sum_{st} \theta_{st} \sum_{uv} d_u d_v q_{st}^{uv} \biggr],
\label{eq:term1}
\end{align}
where we have made use of $\sum_{st} q_{st}^{uv} = 1$ and $\sum_v a_{uv} =
d_u$.
\end{widetext}
The first two terms in~\eqref{eq:term1} are constant for any given network
and hence can be neglected---they are irrelevant for comparing the
likelihood values between different runs on the same network.  The final
term can be rewritten using Eq.~\eqref{eq:appthetagamma} as
\begin{equation}
\sum_{st} \theta_{st} \sum_{uv} d_u d_v q_{st}^{uv}
  = \sum_{st} \sum_{uv} a_{uv} q_{st}^{uv} = \sum_{uv} a_{uv},
\end{equation}
which is also a constant and can be neglected.  Thus only the third term
in~\eqref{eq:term1} need be carried over.

The second sum in~\eqref{eq:ll1} is
\begin{align}
\sum_{\vec{s}} q(\vec{s}) \sum_u \log \gamma_{s_u,x_u}
  &= \sum_{su} q_s^u \log \gamma_{s,x_u} \nonumber\\
  &\hspace{-8em}{} = \sum_{su} q_s^u \sum_x \delta_{x,x_u} \log \gamma_{sx}
   = \sum_{usx} \delta_{x,x_u} \gamma_{sx} \log \gamma_{sx} \nonumber\\
  &\hspace{-8em}{} = \sum_{su} \gamma_{s,x_u} \log \gamma_{s,x_u},
\label{eq:term2}
\end{align}
where we have used Eq.~\eqref{eq:appthetagamma} again in the second line.

The final sum in~\eqref{eq:ll1} is the entropy of the posterior
distribution~$q(\vec{s})$, which is harder to calculate because it requires
not just the marginals of~$q$ but the entire distribution.  We get around
this by making the so-called Bethe approximation~\cite{YFW03}
\begin{equation}
q(\vec{s}) = {\prod_{u<v} \bigl[ q_{s_u,s_v}^{uv} \bigr]^{a_{uv}}\over
             \prod_u \bigl[ q_s^u \bigr]^{d_u-1}},
\label{eq:bethe}
\end{equation}
which is exact on trees and locally tree-like networks, and is considered
to be a good working approximation on other networks.  Substituting this
form into the entropy term gives
\begin{align}
\sum_{\vec{s}} q(\vec{s}) \log q(\vec{s})
 &= \half \sum_{uv} a_{uv} \sum_{st} q_{st}^{uv} \log q_{st}^{uv} \nonumber\\
 &\qquad{} - \sum_u (d_u-1) \sum_s q_s^u \log q_s^u.
\label{eq:term3}
\end{align}

\begin{widetext}
Combining Eqs.~\eqref{eq:term1} to~\eqref{eq:term3} and substituting into
Eq.~\eqref{eq:ll1}, our final expression for the log-likelihood, neglecting
constants, is
\begin{align}
\log P(\mat{A}|\boldsymbol{\Theta},\boldsymbol{\Gamma},\vec{x})
  &= \half \sum_{st} \log \theta_{st} \sum_{uv} a_{uv} q_{st}^{uv}
     + \sum_u \sum_s \gamma_{s,x_u} \log \gamma_{s,x_u}
     - \half \sum_{uv} a_{uv} \sum_{st} q_{st}^{uv} \log q_{st}^{uv}
     \nonumber\\
 &\hspace{20em}{} + \sum_u (d_u-1) \sum_s q_s^u \log q_s^u.
\label{eq:ll2}
\end{align}
The run that returns the largest value of this quantity is the run with the
highest likelihood and hence the best fit to the model.
\end{widetext}

\subsection{Ordered metadata}
\label{sec:ordered}
The case of ordered metadata, such as the body masses used in the food web
example of Fig.~4 in the main paper, is more involved.  Let $P(s|x)$ be the
prior probability that a node belongs to community~$s$ given metadata~$x$.
In most cases the metadata have a finite range and for convenience we
normalize them to fall in the range~$x\in[0,1]$.  (In the rarer case of
metadata with infinite range a transformation can be applied first to bring
them into a finite range.)  One immediate question that arises is what
limitations should be placed on the form of the probability~$P(s|x)$.  We
cannot allow it to take any functional form, such as ones that vary
arbitrarily rapidly, for (at least) two reasons.  First, it would be
unphysical---there are good reasons in most cases to believe that nodes
with infinitesimally different metadata~$x$ have only infinitesimally
different probabilities of falling in a particular group.  In other words,
$P(s|x)$~should be smooth and slowly varying in some sense.  Second, a
function that can vary arbitrarily rapidly can have arbitrarily many
degrees of freedom, which would lead to overfitting of the model.

To avoid of these problems, we enforce a slowly varying prior by writing
the function~$P(s|x)$ as an expansion in a finite set of suitably chosen
basis functions.  In our work we use polynomials of finite degree.  There
is an interesting model selection problem inherent in the choice of degree
which we do not tackle here but which would be a good topic for future
research.

For representing probability functions in $[0,1]$, as here, a convenient
choice of polynomial basis is the Bernstein polynomials of degree~$N$:
\begin{equation}
B_j(x) = {N\choose j} x^j (1-x)^{N-j},\quad j = 0\ldots N.
\end{equation}
Bernstein polynomials have three particular properties that make them
useful for representing probabilities:
\begin{enumerate}
\item They form a complete basis set for polynomials of degree~$N$.
\item They fall in the range $0\le B_j(x)\le1$ for all $x\in[0,1]$ and
  all~$j$.
\item They satisfy the sum rule
\begin{equation}
\sum_{j=0}^N B_j(x) = 1
\label{eq:bernsteinsum}
\end{equation}
for all~$x\in[0,1]$.
\end{enumerate}
The first of these implies that any degree-$N$ representation of the
probability~$P(s|x)$ can be written in the form
\begin{equation}
P(s|x) = \sum_{j=0}^N \gamma_{sj} B_j(x)
\label{eq:contprior}
\end{equation}
for some choice of coefficients~$\gamma_{sj}$.  Moreover, if
$\gamma_{sj}\in[0,1]$ for all $s,j$ then $P(s|x)\in[0,1]$ for all
$x\in[0,1]$, meaning it is a well-defined probability within this domain.
To see this observe first that $P(s|x)\ge0$ when $\gamma_{sj}\ge0$ since
all $B_j(x)\ge0$, and second that for $\gamma_{sj}\le1$ we have
\begin{equation}
P(s|x) = \sum_{j=0}^N \gamma_{sj} B_j(x) \le \sum_{j=0}^N B_j(x) = 1,
\end{equation}
where we have made use of Eq.~\eqref{eq:bernsteinsum}.

Finally, the normalization condition that $\sum_s P(s|x)=1$ for all~$x$ can
be satisfied by requiring that
\begin{equation}
\sum_s \gamma_{sj} = 1
\label{eq:contsum}
\end{equation}
so that
\begin{equation}
\sum_s P(s|x) = \sum_s \sum_{j=0}^N \gamma_{sj} B_j(x)
  = \sum_{j=0}^N B_j(x) = 1.
\end{equation}

We now employ the form~\eqref{eq:contprior} to represent the prior
probabilities in our EM algorithm, writing
\begin{equation}
P(\vec{s}|\boldsymbol{\Gamma},\vec{x}) = \prod_u P(s_u|x_u).
\end{equation}
The only change to the algorithm from the previous case arises when we
maximize the right-hand side of Eq.~\eqref{eq:ineq}.  Instead of maximizing
with respect to the prior probabilities directly, we now maximize with
respect to the coefficients~$\gamma_{sx}$ of the expansion.  The optimal
values of the coefficients are given by
\begin{equation}
\gamma_{sj} = \argmax_{\set{\gamma_{sj}}} \sum_{ut} q_t^u \log \sum_k
              \gamma_{tk} B_k(x_u),
\label{eq:contsolve}
\end{equation}
subject to the constraint~\eqref{eq:contsum}.  One can derive conditions
for the maximum by direct differentiation, but the equations do not have a
closed-form solution, so instead we once again employ Jensen's
inequality~\eqref{eq:jensen} to write
\begin{equation}
\sum_{ut} q_t^u \log \sum_k \gamma_{tk} B_k(x_u)
  \ge \sum_{ut} q_t^u \sum_k Q_k^{tu} \log {\gamma_{tk} B_k(x_u)\over
                Q_k^{tu}},
\label{eq:ineq2}
\end{equation}
which is true for any $Q_j^{su}$ satisfying $\sum_j Q_j^{su} = 1$ for
all~$u,s$.  The exact equality is achieved when
\begin{equation}
Q_j^{su} = {\gamma_{sj} B_j(x_u)\over\sum_k \gamma_{sk} B_k(x_u)},
\label{eq:Qj}
\end{equation}
and the maximum of Eq.~\eqref{eq:contsolve} can be computed by first
maximizing over~$Q_j^{su}$ in this way and then over~$\gamma_{sj}$.  This
suggests an iterative algorithm analogous to the EM algorithm in which one
computes the $Q_j^{su}$ from~\eqref{eq:Qj} and then, using those values,
computes the maximum with respect to~$\gamma_{sj}$ by differentiating the
right-hand side of~\eqref{eq:ineq2} subject to the
condition~\eqref{eq:contsum}, which gives
\begin{equation}
\gamma_{sj} = {\sum_u q_s^u Q_j^{su}\over\sum_{tu} q_t^u Q_j^{tu}}.
\label{eq:contgamma}
\end{equation}
Iterating \eqref{eq:Qj} and~\eqref{eq:contgamma} alternately to convergence
now gives us the coefficients~$\gamma_{sj}$ of the optimal degree-$N$
polynomial prior.  Note that~\eqref{eq:contgamma} always
gives~$\gamma_{sj}$ in the range from zero to one, so that, as discussed
above, the resulting prior~$P(s|x)$ also lies between zero and one and is
thus a lawful probability.

\subsection{Implementation}
The calculations for this paper were implemented in the C programming
language for speed.  There are a number of additional techniques that can
be used to improve speed and convergence.  We find that the majority of the
running time of the algorithm is taken up by the belief propagation
calculations, and this time can be shortened by noting that highly
converged values of the beliefs are pointless in early steps of the EM
algorithm.  The parameter values used to calculate the beliefs in these
steps are, presumably, highly inaccurate since the EM algorithm has not
converged yet, so there is little point spending a large amount of time
waiting for the beliefs to converge to many decimal places when there are
much bigger sources of error in the calculation.  In the calculations of
this paper, we limited the belief propagation to no more than 20 steps at
any point.  In the early stages of the EM algorithm this gives rather crude
values for the beliefs, but these values would not be particularly good
under any circumstances, no matter how many steps we used, because of the
poor parameter values.  In the later stages of the EM algorithm, 20 steps
are enough to ensure good convergence (and indeed we often get good
convergence after many fewer steps than this).

We also place a limit on the total number of iterations of the EM
algorithm, discarding results that fail to converge within the allotted
time.  In the calculations in this paper, this second limit was set at
either 20 or 100 steps.  We have performed some runs with higher limits (up
to 1000 EM steps) but, paradoxically, we find this often gives poorer
results, for instance in our tests on synthetic networks.  This seems to be
because the EM algorithm sometimes converges (as we have said) to the wrong
solution and empirically when it does so it also often converges more
slowly.  By discarding runs that converge slowly, therefore, we tend to
discard incorrect solutions and improve the average quality of our results.

\section{Normalized mutual information}
\label{app:nmi}
In this appendix we discuss the definition of the normalized mutual
information that we use to measure the quality of the results given by our
algorithm.

The most widely used measure of agreement between community divisions and
``ground truth'' variables is the \defn{normalized mutual information}
(NMI), first employed in this context by Danon~\etal~\cite{DDDA05}.  Given
a community division represented by the $n$-element vector~$\vec{s}$ and
discrete metadata represented by~$\vec{x}$, the \defn{conditional entropy}
of the community division is~\cite{CT06}
\begin{equation}
H(\vec{s}|\vec{x}) = -\sum_x P(x) \sum_s P(s|x) \log P(s|x),
\end{equation}
where $P(x)$ is the fraction of nodes with metadata~$x$ and $P(s|x)$ is the
probability that a node belongs to community~$s$ if it has metadata~$x$.
Traditionally the logarithm is taken in base~2, in which case the units of
conditional entropy are bits.

In our case we already know the value of $P(s|x)$: it is equal to the prior
probability~$\gamma_{sx}$ of belonging to community~$s$, one of the outputs
of our algorithm.  Hence
\begin{align}
H(\vec{s}|\vec{x}) &= -\sum_x P(x) \sum_s \gamma_{sx} \log \gamma_{sx}
    \nonumber\\
 &= - \sum_x {n(x)\over n} \sum_s \gamma_{sx} \log \gamma_{sx} \nonumber\\
 &= - {1\over n} \sum_{su} \gamma_{s,x_u} \log \gamma_{s,x_u},
\end{align}
where $n(x) = nP(x)$ is the number of nodes with metadata~$x$ and $n$ is
the total number of nodes in the network, as previously.

The conditional entropy is equal to the amount of additional information
one would need, on top of the metadata themselves, in order to specify the
community membership of every node in the network.  If the metadata are
perfectly correlated with the communities, so that knowing the metadata
tells us the community of every node, then the conditional entropy is zero.
Conversely, if the metadata are worthless, telling us nothing at all about
community membership, then the conditional entropy takes its maximum value,
equal to the total entropy of the community assignment $H(\vec{s}) = -
\sum_s P(s) \log P(s)$.  Alternatively, if we want a measure that increases
(rather than decreases) with the amount of information the metadata give
us, we can subtract $H(\vec{s}|\vec{x})$ from~$H(\vec{s})$, which gives the
(unnormalized) mutual information
\begin{equation}
I(\vec{s}\,;\vec{x}) = H(\vec{s}) - H(\vec{s}|\vec{x}).
\label{eq:mutual}
\end{equation}
This has a range from zero to~$H(\vec{s})$, making it potentially hard to
interpret, so commonly one normalizes it, creating the normalized mutual
information.  There are several different normalizations in use.  As
discussed by McDaid~\etal~\cite{MGH11}, it is mathematically reasonable to
normalize by the larger, the smaller, or the mean of the entropies
$H(\vec{s})$ and $H(\vec{x})$ of the communities and metadata.
Danon~\etal~\cite{DDDA05} in their original work used the mean, while
Hric~\etal~\cite{HDF14}, in their work on lack of correlation between
communities and metadata, used the maximum.  In the present case, however,
we contend that the best choice is the minimum.

Since the maximum value of the mutual information is~$H(\vec{s})$, this
sets the scale on which it should be considered large or small.  Thus one
might imagine the correct normalization would be achieved by simply
dividing $I(\vec{s}\,;\vec{x})$ by~$H(\vec{s})$, yielding a value that runs
from zero to one.  This, however, would give a quantity that was asymmetric
with respect to~$\vec{s}$ and~$\vec{x}$---if the values of the two vectors
were reversed the value of the mutual information would change.  Mutual
information, by convention, is symmetric and we would prefer a symmetric
normalization scheme.  Dividing by $\min[H(\vec{s}),H(\vec{x})]$ achieves
this.  In all the examples we consider, the number of communities is less
than the number of metadata values, in some cases by a wide margin.
Assuming the values of both to be reasonably broadly distributed, this
implies that the entropy $H(\vec{s})$ of the communities will be smaller
than that of the metadata~$H(\vec{x})$ and hence, normally,
$\min[H(\vec{s}),H(\vec{x})] = H(\vec{s})$.  Thus if we define
\begin{equation}
\textrm{NMI} = {I(\vec{s}\,;\vec{x})\over\min[H(\vec{s}),H(\vec{x})]},
\label{eq:nmi}
\end{equation}
we ensure that the normalized mutual information lies between zero and one,
that it has a symmetric definition with respect to $\vec{s}$ and~$\vec{x}$,
and that it will achieve its maximum value of one when the metadata
perfectly predict the community membership.  Other definitions, normalized
using the mean or maximum of the two entropies, satisfy the first two of
these three conditions but not the third, giving values smaller than one by
an unpredictable margin even when the metadata perfectly predict the
communities.

We use the definition~\eqref{eq:nmi} in all the calculations presented in
this paper.

\section{Further examples}
\label{app:examples}
In this appendix we present a number of additional applications of our
methods as well as some additional details on examples described in the
main text.  Summary statistics on all the networks studied are given in
Table~\ref{table:networks}.

\begin{table*}
\centering
\begin{tabular*}{\hsize}{@{\extracolsep{\fill}}llrrlll}
& & & & & \multicolumn{2}{l}{normalized mutual information} \\
network & domain & nodes $n$ & edges $m$ & metadata & blind & with metadata \\
\hline
\multirow{3}{*}{School friendships} & \multirow{3}{*}{social} & \multirow{3}{*}{795} & \multirow{3}{*}{2072} & grade & 0.105--0.384 & 0.881 \\
 & & & & ethnicity & 0.120--0.239 & 0.820 \\
 & & & & gender & 0.000--0.010 & 0.003 \\ 
\multirow{2}{*}{Predator-prey interactions} & \multirow{2}{*}{ecological} & \multirow{2}{*}{488} & \multirow{2}{*}{15\,880} & species body mass & -- & -- \\ 
 & & & & ecological role & 0.348--0.443 & 0.595 \\ 
Internet graph & technological & 46\,676 & 262\,953 & country & 0.396--0.626 & 0.870 \\ 
\multirow{2}{*}{Facebook friendships} & \multirow{2}{*}{social} & \multirow{2}{*}{15\,126} & \multirow{2}{*}{1\,649\,234} & graduation year & 0.573--0.641 & 0.668 \\
 & & & & dormitory & 0.074--0.224 & 0.255 \\ 
Malaria gene recombinations & biological & 297 & 2\,684 & Cys-PoLV labels & 0.077--0.675 & 0.596 \\\hline
\end{tabular*}
\caption{Summaries of real-world network examples and
  their node metadata variables. For each we report the normalized
  mutual information (NMI) between the metadata and the communities
  found without metadata (``blind'') and found using the methods
  described in this paper.}
\label{table:networks}
\end{table*}

\subsection{Facebook friendship network}
The FB100 data set of Traud~\etal~\cite{TMP12} is a set of friendship
networks among college students at US universities compiled from friend
relations on the social networking website Facebook.  The networks date
from the early days of Facebook when its services were available only to
college students and each university formed a separate and unconnected
subgraph in the larger network.  The nodes in these networks represent the
students, the edges represent friend relations on Facebook, and in addition
to the network structure there are metadata of several types, including
gender, college year (i.e.,~year of college graduation), major
(i.e.,~principal subject of study, if known), and a numerical code
indicating which dorm they lived~in.

\begin{figure}
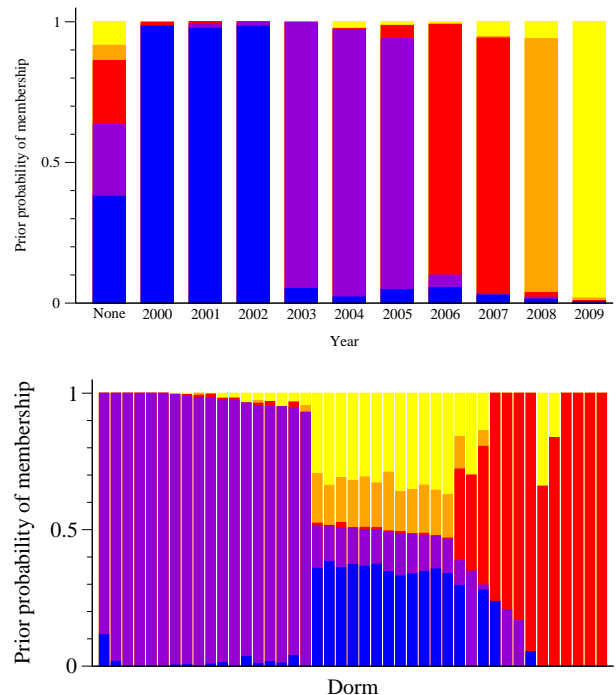

\begin{center}
\includegraphics[width=8cm,clip=true]{harvard_5years.eps}\\
{\ }\\
\includegraphics[width=8cm,clip=true]{harvard_dorms.eps}
\end{center}
\caption{Learned prior probability of community membership for two five-way
  divisions of the Facebook friendship network of Harvard students
  described in the text.  The horizontal axis is (top)~year of graduation
  and (bottom)~dorm, and the colors represent the prior probabilities of
  membership in each of the communities.}
\label{fig:harvard}
\end{figure}

The primary divisions in these networks appear to be by age, or more
specifically by college year.  For instance, we have looked in some detail
at the data for Harvard University, which was the birthplace of Facebook
and its biggest institutional participant at the time the data were
gathered, with $15\,126$ students in the network, spanning college years
2003 to 2009.  There are also a small number of Harvard alumni
(i.e.,~former students) in the data set, primarily those recently
graduated---graduation years 2000--2002.  The top panel in
Fig.~\ref{fig:harvard} shows results from a five-way division of the
network using our algorithm with year as metadata.  Year, for the purposes
of this calculation, was treated as an unordered variable, placing no
constraints on the value of the prior probabilities of community membership
for adjacent years.  One could have treated it as an ordered variable,
which would have constrained adjacent years to have similar priors, but we
did not do that here.  Nonetheless, as we will see, the algorithm finds
communities in which adjacent years tend to be grouped together.

This network provides a good example of the usefulness of the learned
priors in shedding light on the structure of the network.  The figure shows
a visualization of the priors as a function of year, with the colors
showing the relative probability of belonging to each of the communities.
Each of the bars in the plot has the same height of~1 since the prior
probabilities are required to sum to~1, while the balance of colors shows
the distribution over communities.  Examination of the top panel in the
figure shows clearly a division of the network along age lines.  Two
groups, in orange and yellow at the right of the plot, correspond to the
most recent two years of students at the time of the study (graduation
years 2008 and 2009) and the next, in red, accounts for the two years
before that (2006 and 2007).  The purple community corresponds to the next
three years, 2003--2005, while the sixth group, shown in blue, corresponds
to the alumni.  Finally, students for whom year was not recorded are shown
in the column marked ``None,'' which is a mixture of all five groups.

These results align well with the original analysis of the same data by
Traud~\etal~\cite{TMP12}, who performed a traditional community division of
the network and then carried out \textit{post hoc} statistical tests to
measure correlations between communities and metadata.  They found strong
correlations with college year metadata, in agreement with our results.
With the benefit of hindsight the results may appear unsurprising---anyone
who has been to college knows that a large number of your friends are in
the same year as you---but one could certainly formulate competing
hypotheses.  One alternative that Traud~\etal\ considered was that
friendship might be influenced by where students live, with students living
in the same dormitory more likely to be friends, regardless of what year
they are in.  Traud~\etal\ found that there was some evidence for this
hypothesis, but that the effect was weaker than that for age, and our
analysis confirms this.  The bottom panel in Fig.~\ref{fig:harvard} shows a
plot of the priors for a division with dorm as the metadata variable and
there is a clear correlation between dorm and community membership, but it
is not as clean as in the case of age.  There appear to be two groups that
align strongly with particular sets of dorms (colored red and purple in the
figure) while the rest of the dorms are a mix of different communities (the
region in the middle of the figure).  The impression that the community
structure is more closely aligned with graduation year than with dormitory
is also borne out by the normalized mutual information values for the two
divisions.  For the case of graduation year the NMI is 0.668; for dormitory
it is 0.255.

\subsection{Malaria gene recombination network}
Malaria, which is caused by the parasite \textit{P.~falciparum}, is endemic
in tropical regions and is responsible for roughly a million deaths
annually, mostly children in sub-Saharan Africa~\cite{WMR12}.  During
infection, parasites evade the host immune system and prolong the infection
by repeatedly changing a protein camouflage displayed on the surface of an
infected red blood cell. To enable this behavior, each parasite has a
repertoire of roughly 60 immunologically distinct proteins, each of which
is encoded by a \textit{var} gene in the parasite's
genome~\cite{BBKQHKMN05}. These genes undergo frequent recombination,
producing novel proteins by shuffling and splicing substrings from existing
\textit{var} genes.

The process of recombination induces a natural bipartite network with two
types of nodes, \textit{var} genes on the one hand and their constituent
substrings on the other, where each gene node is connected by an edge to
every substring it contains~\cite{LCB13,LCJ14}.  Recombination in these
genes occurs mainly within a number of distinct \textit{highly variable
  regions} (HVRs) and each HVR represents a distinct set of edges among the
same nodes.  Here, we focus on the one-mode gene-gene projections of the
HVR~5 and HVR~6 subnetworks, which have previously been analyzed using
community detection methods without metadata~\cite{LCB13,LCJ14}.  Each of
these one-mode networks consists of 297 genes.

\begin{figure}[t!]
\begin{center}
\subfigure[~Without metadata]{\includegraphics[width=7.8cm]{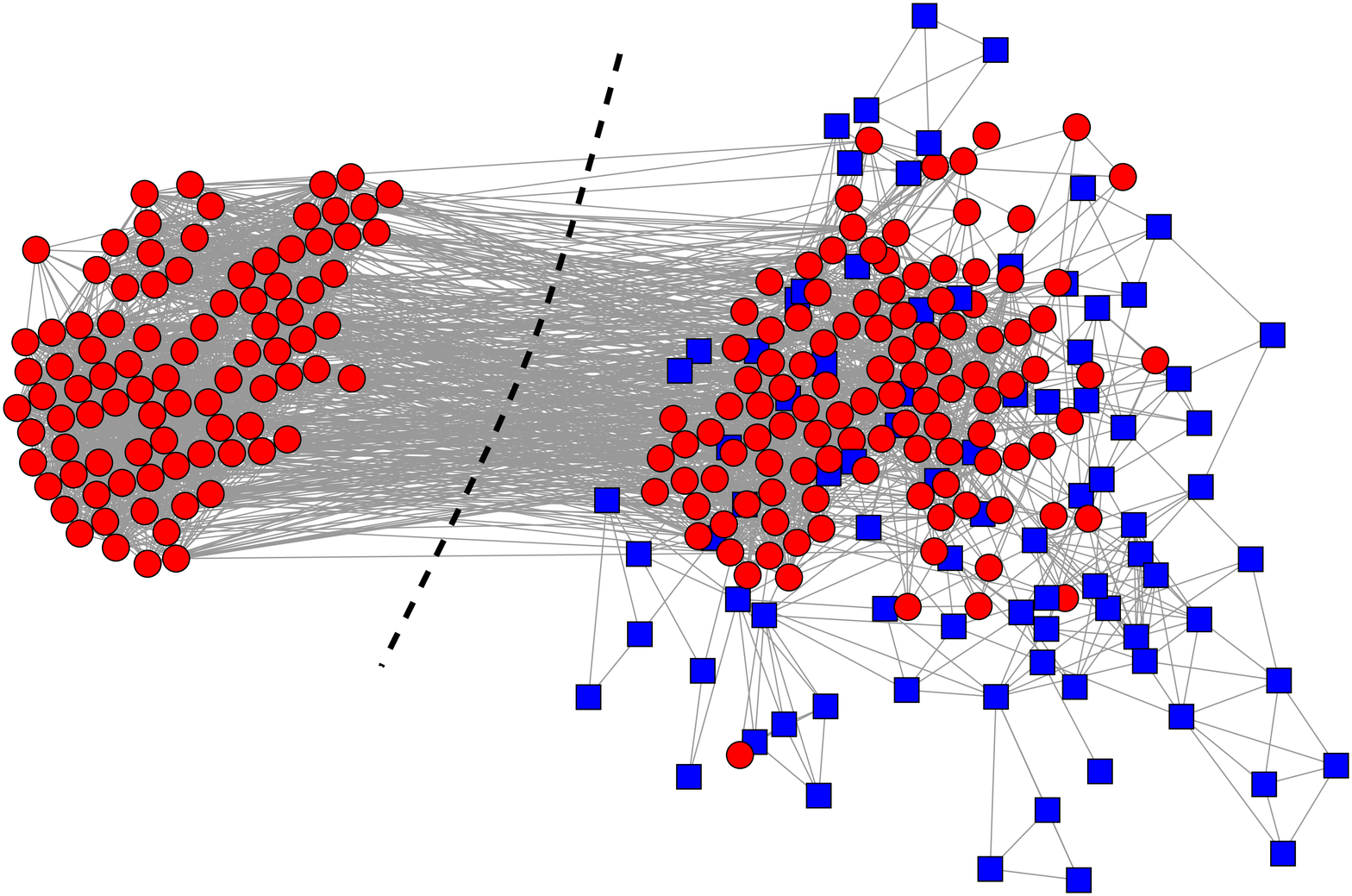}}
\subfigure[~With metadata]{\includegraphics[width=8.4cm]{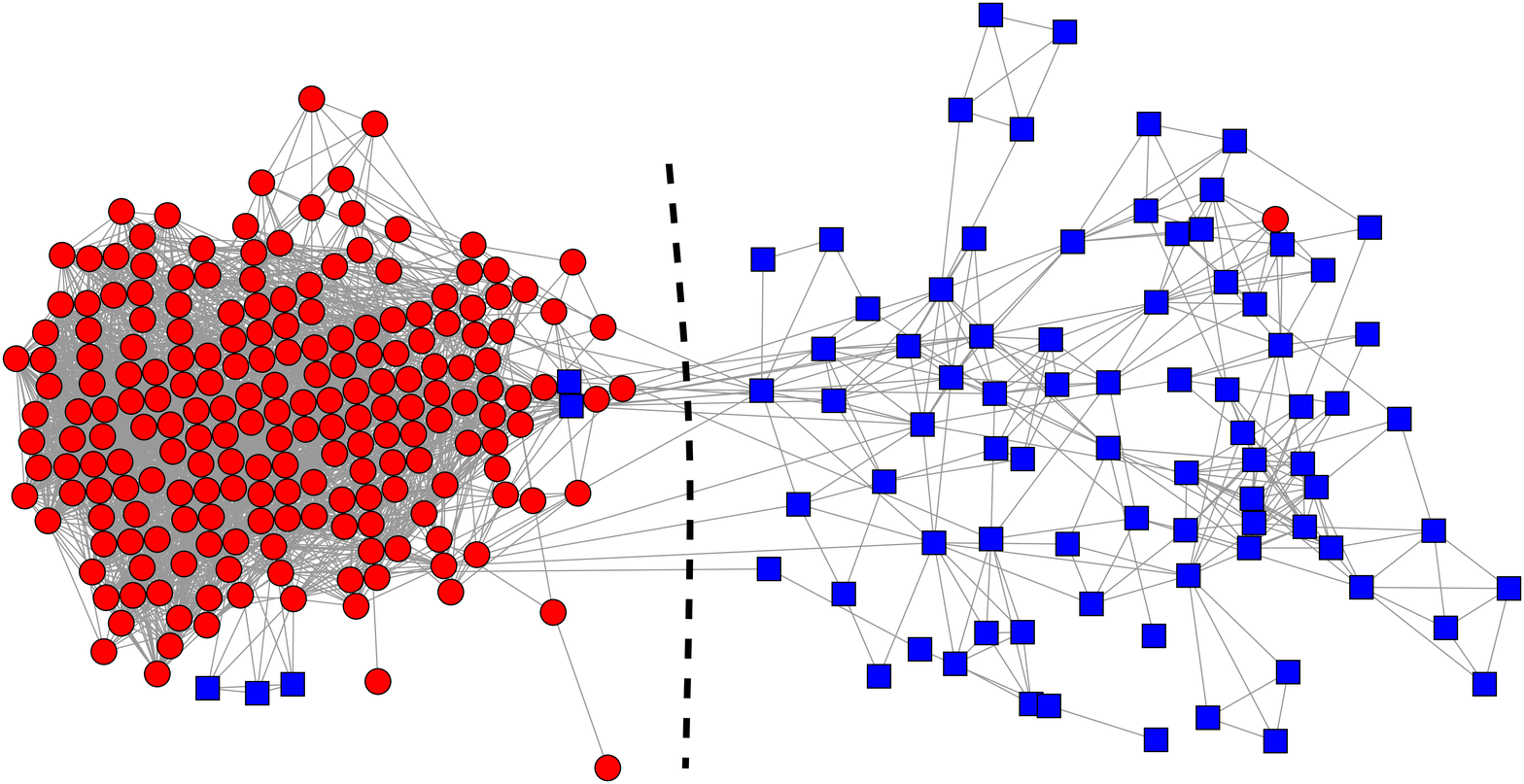}}
\end{center}
\caption{Inferred communities, without metadata and with, for the HVR 6
  gene recombination network of the human malaria parasite
  \textit{P.~falciparum}, where metadata values are the CP labels for the
  genes.  Nodes are colored by their biologically relevant Cys label.}
\label{fig:malaria:HVR6}
\end{figure}

We analyze these networks using the methods described in this paper. As
metadata, we use the Cys labels derived from the HVR~6 sequence and the
Cys-PoLV (CP) labels derived from the sequences adjacent to HVRs~5
and~6~\cite{BBKQHKMN05,WKFMNPBMB09,BKBMKNM07}. Both types of labels depend
only on the sequences' characteristics:\ Cys indicates the number of
cysteines the HVR~6 sequence contains (2~or~4) while CP subdivides the Cys
classifications into 6 groups depending on particular sequence motifs.
Thus, each node has two metadata values, a Cys label and a CP label. The
Cys labels are biologically important because cysteine counts have been
implicated in severe disease phenotypes~\cite{BBKQHKMN05,WKFMNPBMB09}.

\begin{figure}[t!]
\begin{center}
\subfigure[~Without metadata]{\includegraphics[width=7.2cm]{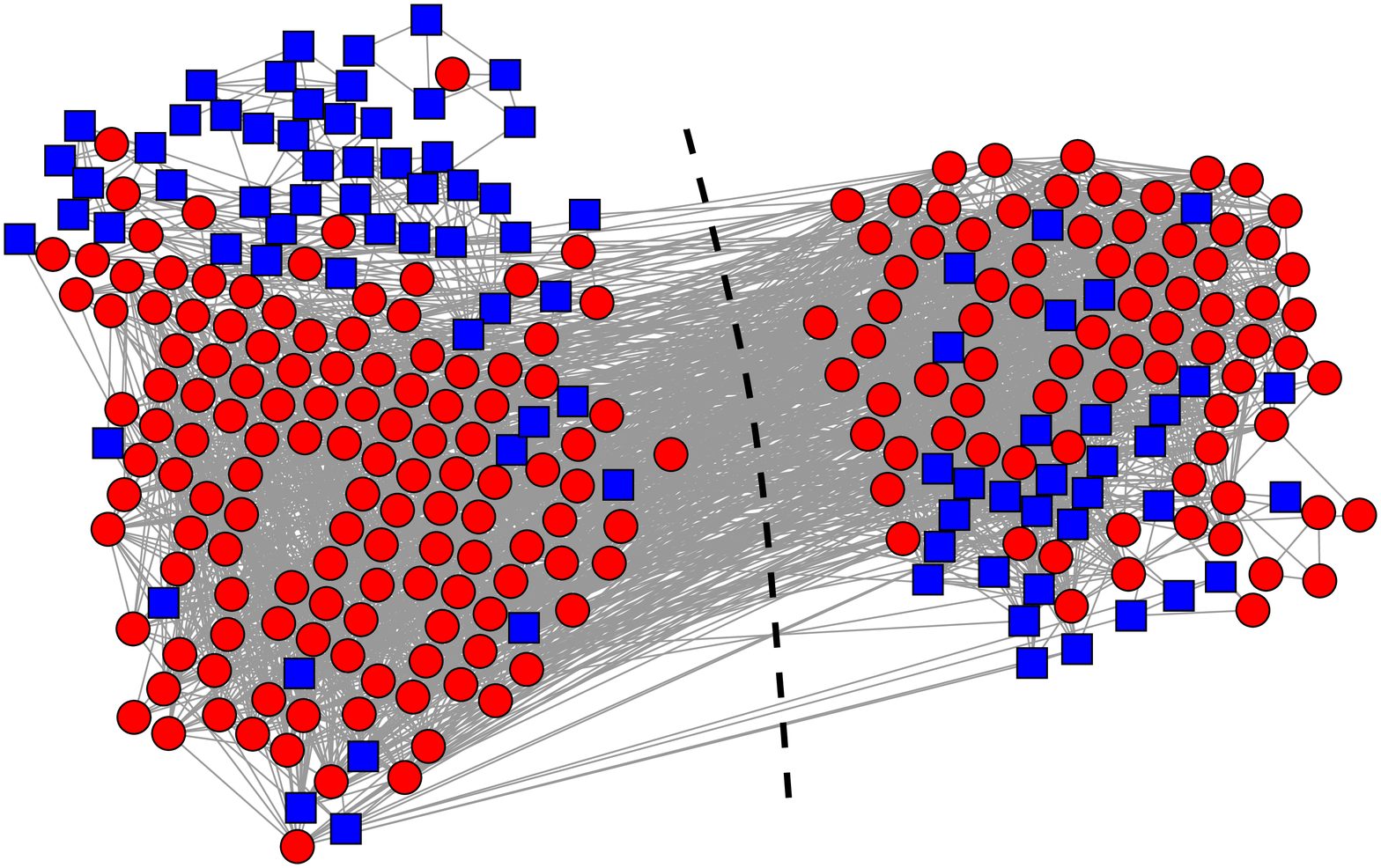}}
\subfigure[~With metadata]{\includegraphics[width=8.2cm]{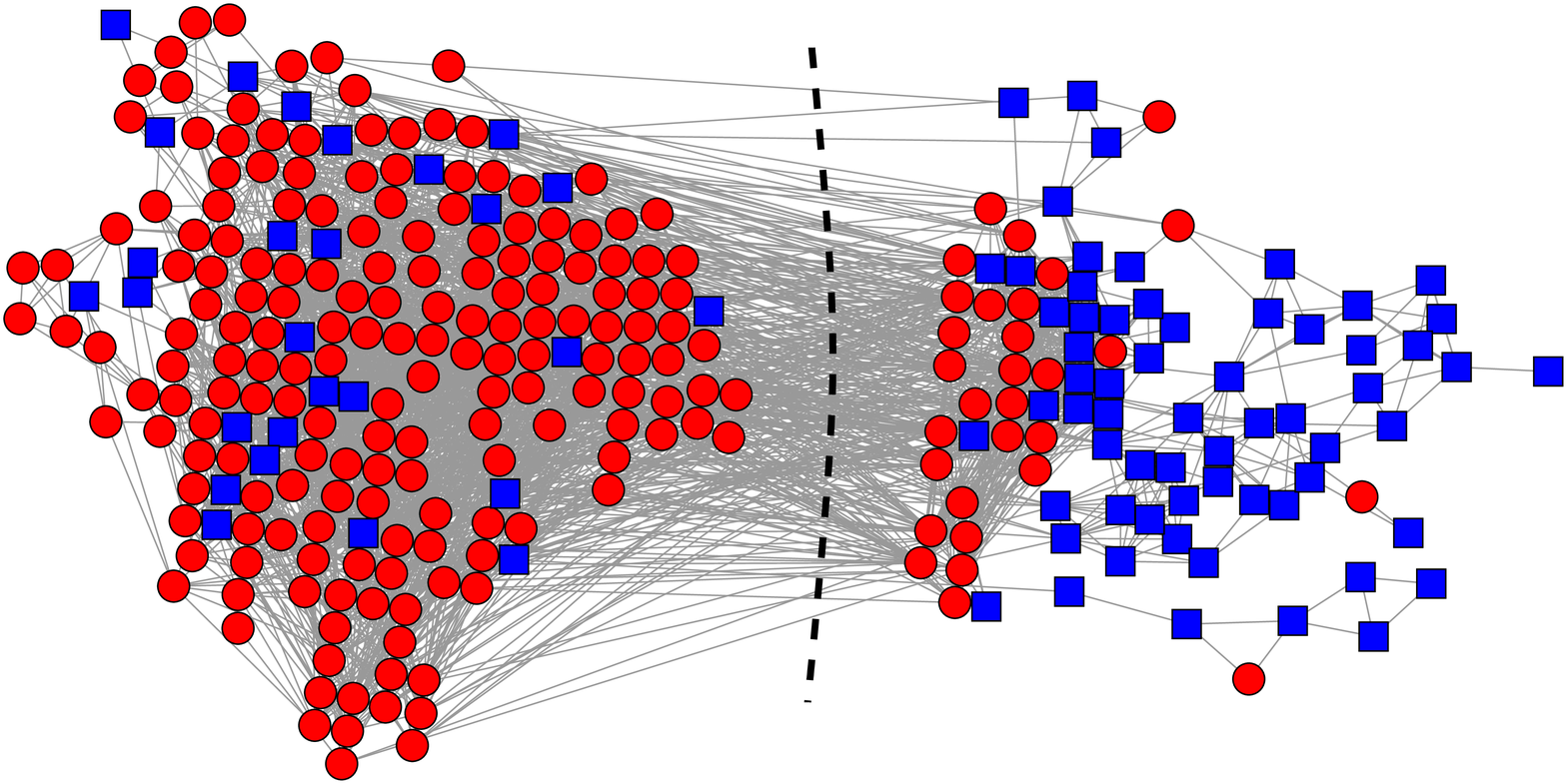}}
\end{center}
\caption{Inferred communities, without metadata and with, for the HVR~5
  gene recombination network of the human malaria parasite
  \textit{P.~falciparum}, where metadata values are the CP labels for the
  HVR~6 network.}
\label{fig:malaria:HVR5}
\end{figure}

In our calculations we use the six CP labels as metadata for a 2-way
community division of the network and then evaluate the degree to which the
inferred communities correlate with the Cys metadata.
Figure~\ref{fig:malaria:HVR6} shows the results for the HVR~6 network with
and without the CP labels as metadata. Without metadata, the Cys labels are
mixed across the inferred groups (Fig.~\ref{fig:malaria:HVR6}a), but with
metadata we obtain a nearly perfect partition
(Fig.~\ref{fig:malaria:HVR6}b).  This indicates that the CP label
correlates well with the network's community structure, a fact that was
obscured in the analysis without metadata.  Furthermore, the inferred
communities correlate strongly with the coarser Cys labels, which were not
shown to the method:\ observing that a gene has two cysteines is highly
predictive (96\%~probability) of that gene being in one group, while having
four cysteines is modestly predictive (67\%~probability) of being in the
other group.  Thus the method has discovered by itself that the motif
sequences that define the CP labels, along with their corresponding network
communities, correlate with cysteine counts and their associated severe
disease phenotypes~\cite{BBKQHKMN05,WKFMNPBMB09}.

The communities in the HVR~6 network represent highly non-random patterns
of recombination, which are thought to indicate functional constraints on
protein structure.  Previous work has conjectured that common constraints
on recombination span distinct HVRs~\cite{LCB13}.  We can test this
hypothesis using the methods described in this paper.  There is no reason
\textit{a priori} to expect that the community structure of HVR~6 should
correlate with that of HVR~5 because the Cys and CP labels are derived from
outside the HVR~5 sequences---Cys labels reflect cysteine counts in HVR~6
while CP labels subdivide Cys labels based on sequence motifs adjacent to,
but outside of, HVR~5.  Applying our methods to HVR~5 without any metadata
(Fig.~\ref{fig:malaria:HVR5}a), we find mixing of the HVR~6 Cys labels
across the HVR~5 communities.  By contrast, using the CP labels as metadata
for the HVR~5 network, our method finds a much cleaner partition
(Fig.~\ref{fig:malaria:HVR5}b), indicating that indeed the HVR~6 Cys labels
correlate with the community structure of HVR~5.

\subsection{Weddell Sea food web}
\begin{figure}
\begin{center}
\includegraphics[width=8cm,clip=true]{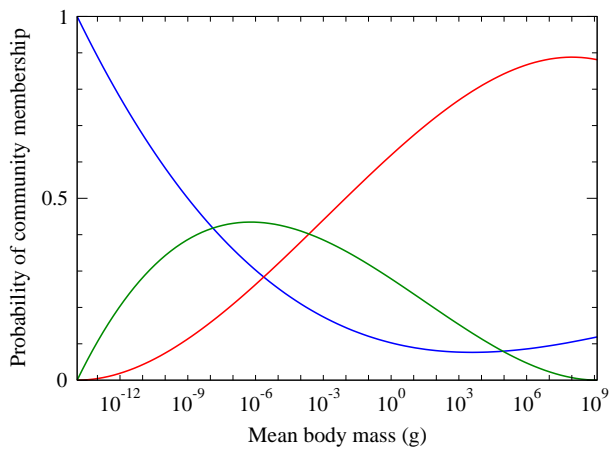}
\end{center}
\caption{Learned priors, as a function of body mass, for the
  three-community division of the Weddell Sea network shown in Fig.~4 of
  the main paper.}
\label{fig:priors}
\end{figure}

As discussed in the main text, the Weddell Sea food web provides an example
of the ``ordered'' metadata type in the body mass of species.  A three-way
community division of the network with the log of species' average body
mass as metadata produces the division shown in Fig.~4 of the paper.  The
prior probabilities as functions of body mass are of interest in their own
right.  They are shown in Fig.~\ref{fig:priors}.  Although, as described in
Section~\ref{sec:ordered} of the paper, the log mass is rescaled in our
calculations to the range $[0,1]$, the horizontal axis in the figure is
calibrated to read in terms of the original mass in grams, so the prior
probabilities of belonging to each of the three communities can be simply
read from the figure.  The blue, green, and red curves correspond
respectively to the communities labeled 1, 2, and~3 in Fig.~4.  Thus a
species with a low mean mass of $10^{-12}\,$g has about an 80\% probability
of being in community~1, a 20\% probability of being in community~2, and
virtually no chance of being in community~3.  Conversely, a species with
mean body mass of $10^8\,$g (which could only be a whale) has about a 90\%
chance of being in community~3, 10\% of being in community~2, and almost no
chance of being in community~1.

\end{document}